\documentclass[11pt,3p,review,authoryear]{elsarticle}

\usepackage{amsmath}
\usepackage{bm}
\usepackage[singlelinecheck=false]{caption}
\usepackage{multirow}
\usepackage{adjustbox}
\usepackage{pdflscape}
\usepackage[colorlinks,allcolors=red]{hyperref}
\usepackage{hhline}
\usepackage{mathtools}

\usepackage{titlesec}
\titleformat{\paragraph}
{\normalfont\normalsize\bfseries}{\theparagraph}{1em}{}
\titlespacing*{\paragraph}
{0pt}{3.25ex plus 1ex minus .2ex}{1.5ex plus .2ex}

\begin{document}

\begin{frontmatter}

\title{Real-time Prediction of the Great Recession and the Covid-19 Recession}

\author[inst1]{Seulki Chung}

\affiliation[inst1]{organization={GSEFM, Department of Empirical Economics, Technische Universität Darmstadt},
            addressline={Karolinenpl.5}, 
            city={Darmstadt},
            postcode={64289}, 
            country={Germany}}



\begin{abstract}
This paper uses standard and penalized logistic regression models to predict the Great Recession and the Covid-19 recession in the US in real time. It examines the predictability of various macroeconomic and financial indicators with respect to the NBER recession indicator. The findings strongly support the use of penalized logistic regression models in recession forecasting. These models, particularly the ridge logistic regression model, outperform the standard logistic regression model in predicting the Great Recession in the US across different forecast horizons. The study also confirms the traditional significance of the term spread as an important recession indicator. However, it acknowledges that the Covid-19 recession remains unpredictable due to the unprecedented nature of the pandemic. The results are validated by creating a recession indicator through principal component analysis (PCA) on selected variables, which strongly correlates with the NBER recession indicator and is less affected by publication lags.
\linebreak
\end{abstract}


\end{frontmatter}
\newpage
\section{Introduction}

Business cycles are characterized by alternating expansion and contraction phases in aggregate economic activity. The precise dating of business cycle turning points has been a primary concern for policymakers and businesses, as it enables them to make informed decisions. However, this dating process is time-consuming and typically takes months or even years to determine the exact date of business cycle turning points that already lie in the past. Therefore, the primary objective of related research has been to measure the business cycle and then develop a modeling framework that can accurately predict the phases of business cycles. In particular, there has been more interest in predicting the phase of economic contraction, often referred to as a recession, as it is often accompanied by severe economic downturns. Recessions are not triggered by a single shock, which makes them difficult to predict. The Great Recession and the Covid-19 recession are the two most recent recessions in the US and can be safely assumed to exhibit their own unique characteristics: The Great Recession is mainly caused by the collapse of financial systems, whereas the Covid-19 recession has its roots in the widespread pandemic. This paper aims to address two interrelated research questions: First, it investigates whether these two recessions of different natures can be predicted in real time. Second, it finds out which factors contribute to their predictability. 

Earlier work by \citet{burns1946} is among the first to provide a comprehensive summary of how to measure business cycles and identify their turning points. \citet{stock1989, stock1993} develop a modeling framework to measure business cycles by extracting an alternative index based on the co-movements in various components of economic activity. A common approach, based on their work, to measure business cycles involves using indicators that well represent the overall economic activity of a country, such as real GDP and the industrial production index, as reference series. The focus is on the deviation from trends or cyclical fluctuations around the trend line. While real GDP may better represent the overall economic activity of a country, the industrial production index has the advantage of monthly data availability. In this case, predicting business cycles can take the form of modeling and forecasting GDP growth, for instance. On the other hand, the reference series can be transformed into a binary variable marking two possible states of the economy: boom or recession. This transformation often follows the approach of \citet{bry1971}.

Earlier studies on forecasting binary economic events focused on applying logit and probit models and Markov regime-switching models. \citet{canova1994} employs probit models to investigate the predictability of financial crises in the US in the pre-World War I era. \citet{dueker1997} extends the static model and adds dynamics by incorporating a lagged recession indicator in the specification of the probit model. \citet{chauvet2006} use a Markov regime-switching model and produce new measures to identify business cycle turning points. \citet{tian2019} provide extensions to the application of Markovian models by incorporating the temporal autocorrelation of binary recession indicators, thus predicting US recessions. Newer studies actively apply machine learning techniques to forecasting business cycle turning points. \citet{berge2015} employs the Bayesian model averaging technique and boosting algorithms to investigate US business cycle turning points. \citet{dopke2017} use boosted regression trees to predict German economic recessions. \citet{davig2019} apply the naive Bayes model to recession forecasting, showing that their model outperforms logistic regression models. \citet{chatzis2018} and \citet{vrontos2021} exploit various machine learning techniques to forecast stock market crises and US recessions, respectively. 

This paper applies two types of binary choice models: the standard logistic regression models and the penalized logistic regression models for recession forecasting. The models include the basic logit model, the weighted logit model, the LASSO logit model, the Ridge logit model, and the Elastic Net logit model. Except for the basic logit model, a different weighting scheme is employed for two different classes of the binary recession indicator to account for class imbalance. 

Many macroeconomic variables frequently used as predictors for recession forecasting are subject to multiple revisions. It is recognized that the usage of the latest available data, rather than real-time data, can affect the forecasts and distort the forecast evaluation results (\citet{stark2002}). Hence, this paper utilizes a real-time dataset to assess the real-time predictability of recessions. 

Forecasts are generated for five different forecast horizons: nowcasting, immediate-term (1-month-ahead), short-term (3-month-ahead), medium-term (6-months-ahead), and long-term (12-months-ahead) setup. Two sets of metrics are used to evaluate forecast performance. The first set is threshold-invariant and measures performance based only on predicted probabilities. The second set depends on the cutpoint that turns predicted probabilities into binary outcomes. These thresholds are endogenously optimized during the cross-validation process and thus vary over time. 

The LASSO models conduct automatic variable selections during model estimation. Based on the LASSO logit models, five predictors that are not left out during at least 80\% of the two recession periods are chosen and plotted along with their coefficients and predicted recession probabilities to explain the recessions' predictability compared to each other.

The main findings can be summarized as follows: the penalized logit models exhibit superior performance compared to the standard logit models. According to their forecasts, the Great Recession is largely predictable in real time with high accuracy, whereas the Covid-19 recession remains unpredictable to any extent. Furthermore, the empirical study confirms that the term spread remains the foremost economic indicator for predicting the Great Recession. The Great Recession, rooted in the financial system's collapse, resembles past recessions influenced by high inflation and tight monetary policy. The model recognizes this kind of recession and reacts to changes in the term spread by adjusting the coefficients. In contrast, the Covid-19 recession, triggered by the pandemic and economic shutdown, is unpredictable beforehand, and the model adjusts to the negative term spread differently, reducing the effect of that variable. It becomes evident that major macroeconomic and financial indicators can predict recessions of an economic and financial nature. However, they cannot forecast a recession caused by an exogenous event such as a pandemic.

This paper is structured as follows. Section 2 provides a detailed description of the data used in the study. Section 3 introduces the models and forecast evaluation metrics employed. Section 4 outlines the research methodology and presents the main results. Section 5 conducts robustness checks to validate the findings. Finally, Section 6 concludes the paper by summarizing the key findings and their implications.

\section{Real-time data}

Previous research on the identification and prediction of business cycle turning points is often based on a set of macroeconomic indicators. However, these macroeconomic variables are mostly estimates subject to multiple revisions afterward. \citet{stark2002} show that the usage of the latest available data rather than real-time data does affect the forecasts. Thus, comparisons between the forecasts generated from new models and benchmark forecasts generated in real time should be based on real-time data. As the primary interest of my research lies in the real-time predictability of both the Great Recession and the Covid-19 recession in the US, it is essential to work with real-time data. Recessions might be ex-post predictable based on the latest available data; however, it was not available before the recessions. I use the same data that was available to the real-time forecaster for out-of-sample forecasting to assess whether the recessions were predictable.

The predictive data consists of 194 real-time vintages in months, covering a range of macroeconomic and financial market variables from February 1967 to October 2021. The real-time vintage for out-of-sample forecasting begins in November 2006. Based on prior research in existing literature, such as \citet{vrontos2021}, and considering the availability of real-time data, I choose a set of predictors that includes 25 macroeconomic and financial variables from different categories such as output, income, prices, the labor market, the housing market, money and credit, and the financial market. A list with detailed descriptions of these variables is presented in Table 1. The data frequency ranges from daily to quarterly. The mean is used to aggregate higher frequency data to monthly data for model estimation. The variables with quarterly frequency are transformed into monthly equivalents using natural cubic spline interpolation: In a month $t$, all available data up to $t$ are fed into the algorithm to find the interpolating cubic spline. Once it is found, data of monthly frequency are generated as the points on the spline curve between the knots.

\renewcommand{\arraystretch}{1}
\begin{table}[!ht]
    \caption{Overview of predictors}
    \resizebox{\textwidth}{!}{\begin{tabular}{llllll}
    \hline
    Nr. & Predictive variable & Abbreviation & Category & Transformation & Frequency \\
    \hline
    1 & Average hourly earnings of production and nonsupervisory employees & AHETPI & Income & Percent change & Monthly\\
    2 & Average weekly hours of production and nonsupervisory employees & AWHNONAG & Labor market & percent change & Monthly \\
    3 & Moody's BAA yield & BAA &Money and credit & First-order difference & Monthly\\
    4 & Moody's BAA yield relative to 10-Year treasury yield & BAA10YM & Money and credit & First-order difference & Monthly \\
    5 & Real manufacturing and trade industries sales & CMRMTSPL & Output & Log growth rate & Monthly \\
    6 & Corporate profits after tax & CP & Income & Log growth rate & Quarterly \\ 
    7 & Real disposable personal income & DSPIC96 & Income &Log growth rate & Monthly \\
    8 & Effective federal funds rate & FEDFUNDS & Financial market & First-order difference & Monthly \\
    9 & Real gross domestic product & GDPC1 & Output & Log growth rate & Quarterly \\
    10 & Privately-owned housing units started & HOUST & Housing market & Log growth rate & Monthly \\
    11 & Industrial production index & INDPRO & Output & Log growth rate & Monthly \\
    12 & Real M1 money stock & M1REAL & Money and credit & First-order difference & Monthly\\
    13 & Real M2 money stock & M2REAL & Money and credit & First-order difference & Monthly\\
    14 & Non-farm payroll total & PAYEMS & Labor market & Log growth rate & Monthly\\
    15 & Real personal consumption expenditures & PCEC96 & Prices & Log growth rate & Monthly \\
    16 & Privately-owned housing units permitted & PERMIT & Housing market & Log growth rate & Monthly \\
    17 & Producer price index by all commodities & PPIACO & Prices & Log growth rate & Monthly \\
    18 & Private residential fixed investment & PRFI &Housing market & Log growth rate & Quarterly \\
    19 & S\&P 500 index & SP500 & Financial market & Log growth rate & Daily \\
    20 & 3-month treasury bill rate  & TB3MS &Financial market & First-order difference & Monthly \\
    21 & Term spread - 5-year treasury yield minus 3-month treasury bill rate & T5Y3MM & Financial market & None & Monthly \\
    22 & Consumer Sentiment - University of Michigan & UMCSENT & Prices & Log growth rate & Monthly \\
    23 & Unemployment rate & UNRATE & Labor market & First-order difference & Monthly \\
    24 & Producer price index by commodity: final demand: finished goods & WPSFD49207 & Prices & Log growth rate & Monthly \\
    25 & Real personal income excluding current transfer receipts & W875RX1 & Income & Log growth rate & Monthly\\
    \hline
    \end{tabular}}
    \caption*{The table presents a list of predictive variables in alphabetical order based on their abbreviations according to ALFRED. It includes information on their respective categories, transformations applied to ensure stationarity, and their original data frequency.}
\end{table}

I retrieve most of the data vintages, monthly snapshots, of these variables from ALFRED. Among the selected predictors, there are a few variables whose real-time vintages prior to 2013 are not available in ALFRED. The first set of those variables includes real personal income excluding transfer receipts and real manufacturing and trade sales. \citet{chauvet2008} use them along with total non-farm payroll employment and industrial production index to identify business cycle dates in real time. Jeremy Piger provides the real-time dataset of these series on his website. His dataset ends in August 2013, but can be easily extended past 2013 through ALFRED. The second set contains real M1 and M2 money stock, whose earliest available real-time vintage in ALFRED dates back to January 2014. Prior to this date, real-time vintages of real M1 and M2 stock are constructed by deflating nominal M1 and M2 stock with the consumer price index, whose real-time vintages are readily available in ALFRED.      

Each series is individually assessed and transformed to be stationary. \citet{park2000} developed an asymptotic theory for time series binary choice models with nonstationary independent variables and proved the maximum likelihood estimator to be consistent. However, the consistency of the maximum likelihood estimator under model specifications other than logit or probit models has not been proven yet. Hence, I take the conservative step of using stationary time series variables on the right-hand side of the regression equation. All data, apart from the binary recession indicator, are seasonally adjusted and standardized. Sometimes, data are published with a time delay or simply have missing values. In these cases, the $k$-nearest-neighbor imputation method is implemented in the data pre-processing stage. The results do not change when I use a different method, such as the bagged tree imputation algorithm, which is often regarded as a superior method in the existing literature but is associated with higher computation costs (\citet{stekhoven2011}).
 
For business cycle dates, I refer to the NBER-defined business cycle expansion and contraction dates to determine US recessions because it is widely regarded as the benchmark for the US business cycle in the existing literature. Specifically, recession months are defined as those from the period following the peak through the trough. The rest of the months are considered booms. The oldest vintage in ALFRED for the NBER recession indicator is from September 2014. The monthly vintages before that date are hand-collected and constructed based on the official announcements of the NBER Business Cycle Dating Committee. 

\begin{table}[!ht]
        \caption{\\US Business Cycle dates} 
    \begin{tabular}{llll}
    \hline
    Date & Type & Duration & Announcement \\
    \hline
    1980:01 & Peak & 6 & 1980:06(+5) \\
    1980:07 & Trough & 12 & 1981:07(+12) \\
    1981:07 & Peak & 16 & 1982:01(+6) \\
    1982:11 & Trough & 92 & 1983:07(+8) \\
    1990:07 & Peak & 8 & 1991:04(+9)\\
    1991:03 & Trough & 120 & 1992:12(+21)\\
    2001:03 & Peak & 8 & 2001:11(+8)\\
    2001:11 & Trough & 73 & 2003:07(+20)\\
    2007:12 & Peak & 18 & 2008:12(+12)\\
    2009:06 & Trough & 128 & 2010:09(+15)\\
    2020:02 & Peak & 2 & 2020:06(+4)\\
    2020:04 & Trough & ongoing & 2021:07(+15)\\
    \hline
    \end{tabular} \\
    \caption*{The table reports NBER business cycle dates, their type, the duration of either contraction or expansion in months, and their announcement time, with publication lags in parentheses, for the period between 1980 and 2021.} 
\end{table} 

One practical issue concerning NBER business cycle dates is that they are often announced with considerable publication lags. Table 2 presents the dates of peaks and troughs of the US business cycle along with announcement dates from 1980 to 2021. The time delay in parentheses varies between 5 and 21 months for the recent six US recessions. In particular, the publication lags of troughs are, on average, about 2 times longer than those of peaks. Once fixed, NBER business cycle dates have not been revised since 1980. However, the publication lags complicate the work of creating real-time vintages of the NBER recession indicator. The NBER recession indicator in ALFRED is built under the assumption that the previous state does not change until the NBER formally announces a new turning point. I follow this approach, although these publication lags may cause biased estimators, especially regarding the troughs. For instance, the end of the Great Recession was announced in September 2010 to be in June 2009. In this time span of 15 months, the recession indicator keeps giving false alarms that the economy is in recession. Hence, as robustness checks in Section 5, I employ two other approaches to mitigate this concern.

\section{Econometric methodology}

\subsection{Model specifications}

This section discusses the technical details of the models used to forecast two recent recessions in the US. Recessions are binary events; thus, standard logit models are estimated first and used as the reference model to compare with other, more advanced methods. Specifically, weighted logit models are considered next to account for class imbalance, as the number of months in booms outweighs the number of months in recessions. Finally, a number of penalized logit models are used to impose shrinkage on the regression coefficients and thus avoid problems of multicollinearity and overfitting.

\subsubsection{Logistic regression model}
Logit models are broadly employed to model the probability of a binary event. In terms of recession forecasting, it means that they can model and predict the probability of a recession based on a wide range of predictor variables. The dependent variable is the recession indicator, taking values of 1 if the economy is in recession and 0 otherwise. Let the outcomes of the recession indicator be denoted by $Y_t$. Then, they are assumed to be realizations of independent Bernoulli random variables with a recession probability of $p_t$. Logit models aim to estimate the conditional probability of a recession $P(Y_t|\bm{X}_{t-h})$ using a linear predictor function that is a linear combination of the explanatory variables. As the values of the linear predictor function can range from $-\infty$ to $+\infty$, the cumulative distribution function of a logistic distribution is used to map the values into the range of 0 to 1. These values can then be interpreted as the recession probability. Formally, the logit model can be specified in the following form:

\begin{align*}
    \ln\Big(\frac{p_t}{1-p_t}\Big)=\bm{x'}_{t-h}\bm{\beta},
\end{align*}
where $p_t=P(Y_t|\bm{X}_{t-h})$ is the probability of recession, and $\bm{x'}_{t-h}$ is the data vector at time $t-h$, including the constant. The parameters $\bm{\beta}$ can be estimated using maximum likelihood methods. The log-likelihood function for a Bernoulli distribution can be written as follows:
\begin{align*}
    \ln\mathcal{L}(Y|\bm{X},\bm{\beta})=\sum_{t=1}^T[y_t\ln p_t+(1-y_t)\ln (1-p_t)].
\end{align*}
Then the parameters estimates are given by
\begin{align*}
    \bm{\hat{\beta}}_{Logit}=\textrm{argmax}_{\bm{\beta}}[\sum_{t=1}^T[y_t\ln p_t+(1-y_t)\ln (1-p_t)]].
\end{align*}
Based on the estimated parameters, the probability of recession can be computed as
\begin{align*}
    \hat{p_t}=\frac{1}{1+e^{-\bm{x'}_{t-h}\bm{\hat{\beta}}}}.
\end{align*}

\subsubsection{Weighted logistic regression model}
Recessions in the US are not rare but certainly less frequent than booms. Booms are, on average, 6 times longer than recessions. For the estimates of the parameters in the standard logit models, class imbalance is not a problem per se. However, \citet{king2001} show that it affects the estimate of the model intercept, biasing all the predictive probabilities. Furthermore, \citet{peduzzi1996} state that the logistic regression coefficients are biased in both positive and negative directions if the number of events per variable is below 10. The number of recession months per variable in my real-time datasets varies between 2.6 and 3.9. Hence, the estimates of the standard logit models are suspected to be biased. One way to address this concern is to use weighted logit models. In this type of model, different weights are assigned to different classes. \citet{manski1977} propose weighted maximum likelihood estimation for the logit models and prove that these estimators are strongly consistent and asymptotically normal. The weighted log-likelihood function adds the weight terms to the previous log-likelihood function as follows:
\begin{align*}
    \ln\mathcal{L}(Y|\bm{X},\bm{\beta})=\sum_{t=1}^T[w^+_t y_t\ln p_t+w^-_t (1-y_t)\ln (1-p_t)],
\end{align*}
where $w^+$ is the weight for the positive class, corresponding to a recession, and $w^-$ for the negative class. The weights are defined as $w^+_t=\frac{1}{N^+}\times\frac{1}{2}$ and $w^-_t=\frac{1}{N^-}\times\frac{1}{2}$ and thus add up to 1. $N$ denotes the total number of months, and $N^+$ and $N^-$ correspond to the number of months in recessions and the number of months in booms, respectively. The class with fewer events gets a higher weight proportional to the class distribution. \footnote{\citet{manski1977} define the class weights as $w^1=\frac{\tau}{\bar y}$ and $w^0=\frac{1-\tau}{1-\bar y}$ where $\tau$ is the fraction of recession months in the population. However, the class distribution in the population is unknown. Thus, for convenience, I resort to the approach solely based on the class distribution in the sample, as it is widely used in the literature.}
Again, the parameters are estimated by 
\begin{align*}
    \bm{\hat{\beta}}_{wLogit}=\textrm{argmax}_{\bm{\beta}}[\sum_{t=1}^T[w^+_t y_t\ln p_t+w^-_t (1-y_t)\ln (1-p_t)]].
\end{align*}

\subsubsection{Ridge logistic regression model}
The Ridge logit model is an extension of the original Ridge regression model proposed by \citet{hoerl1970}. It is a penalized logistic regression that adds a penalty term to the likelihood function to shrink the estimated coefficients toward zero. In the presence of a large predictor set, including the penalty term helps to avoid the overfitting problem and alleviates the issue of multicollinearity. When both the log-likelihood function and the shrinkage penalty term are subject to a summarized maximization problem, the optimal result will be achieved by shrinking the coefficients of only those variables that are highly correlated. The penalty term also keeps the estimates of less important variables close to zero and, in this way, dampens their effects. Specifically, the Ridge model uses a shrinkage penalty that employs the $\textrm{L}^2$-norm of the parameter vector. The maximum likelihood estimator is then obtained by
\begin{align*}
    \bm{\hat{\beta}_{Ridge}}=\textrm{argmax}_{\bm{\beta}}\Big[\ln\mathcal{L}(Y|\bm{X},\bm{\beta})-\lambda\sum_{i=2}^K\beta_i^2\Big],
\end{align*}
where $\lambda > 0$ is a parameter that determines the size of the penalty. A higher $\lambda$ shrinks the coefficients toward zero more strongly. The hyperparameter $\lambda$ is usually determined via cross-validation techniques, which will be discussed in detail in the next section. 

\subsubsection{LASSO logistic regression model}
An alternative approach is the LASSO (Least Absolute Shrinkage and Selection Operator) model, first introduced by \citet{tibshirani1996}. LASSO logit models go one step further and allow for automatic variable selection by setting the coefficients of less important variables to zero. Hence, this modeling approach entirely eliminates a number of predictors. The objective function imposes a shrinkage penalty by adding the $\textrm{L}^1$-norm of the parameter vector to the log-likelihood function and is given by
\begin{align*}
    \bm{\hat{\beta}_{LASSO}}=\textrm{argmax}_{\bm{\beta}}\Big[\ln\mathcal{L}(Y|\bm{X},\bm{\beta})-\lambda\sum_{i=2}^K|\beta_i|\Big].
\end{align*}

\subsubsection{Elastic net logistic regression model}

The Elastic Net model was originally developed by \citet{zou2005} and is a generalization of the previous two models. It combines the shrinkage penalty terms of both LASSO and Ridge logit models. The LASSO part performs automatic variable selection, while the Ridge part stabilizes the estimates. In this way, the Elastic Net model can deal with a set of highly correlated predictors. The parameter estimates are computed by maximizing the following function with respect to the vector of parameters $\bm{\beta}$:
\begin{align*}
    \bm{\hat{\beta}_{EN}}=\textrm{argmax}_{\bm{\beta}}\bigg[\ln\mathcal{L}(Y|\bm{X},\bm{\beta})-\lambda\Big((1-\alpha)\sum_{i=2}^K\beta_i^2+\alpha\sum_{i=2}^K|\beta_i|\Big)\bigg].
\end{align*}
There are now two hyperparameters, $\lambda$ and $\alpha$. Setting $\alpha$ equal to 1 transforms the Elastic Net model to the LASSO model, whereas $\alpha$ equal to 0 changes it to the Ridge model. In the empirical analysis, I employ the cross-validation technique to choose the optimal value of $\alpha$ among a set of values for $\alpha \in (0.25, 0.5, 0.75)$.

\subsection{Empirical design and analysis} 

The estimation of penalized logit models involves the optimization of hyperparameters. \citet{bergmeier2012} support the use of cross-validation techniques for time series evaluation. In particular, they suggest applying a blocked form of cross-validation for time series data due to theoretical concerns that it contains time dependencies. If data are randomly split into $k$ sets, as in the standard $k$-fold cross-validation, then future data may be used to predict the past. Time-evolving effects are then completely ignored. \citet{cerqueira2020} also share their view and provide further evidence that blocked cross-validation can be applied to stationary time series. Hence, I follow their approach and use a modified version of blocked cross-validation.

\begin{figure}[!ht]
\caption{}
\includegraphics[width=\textwidth]{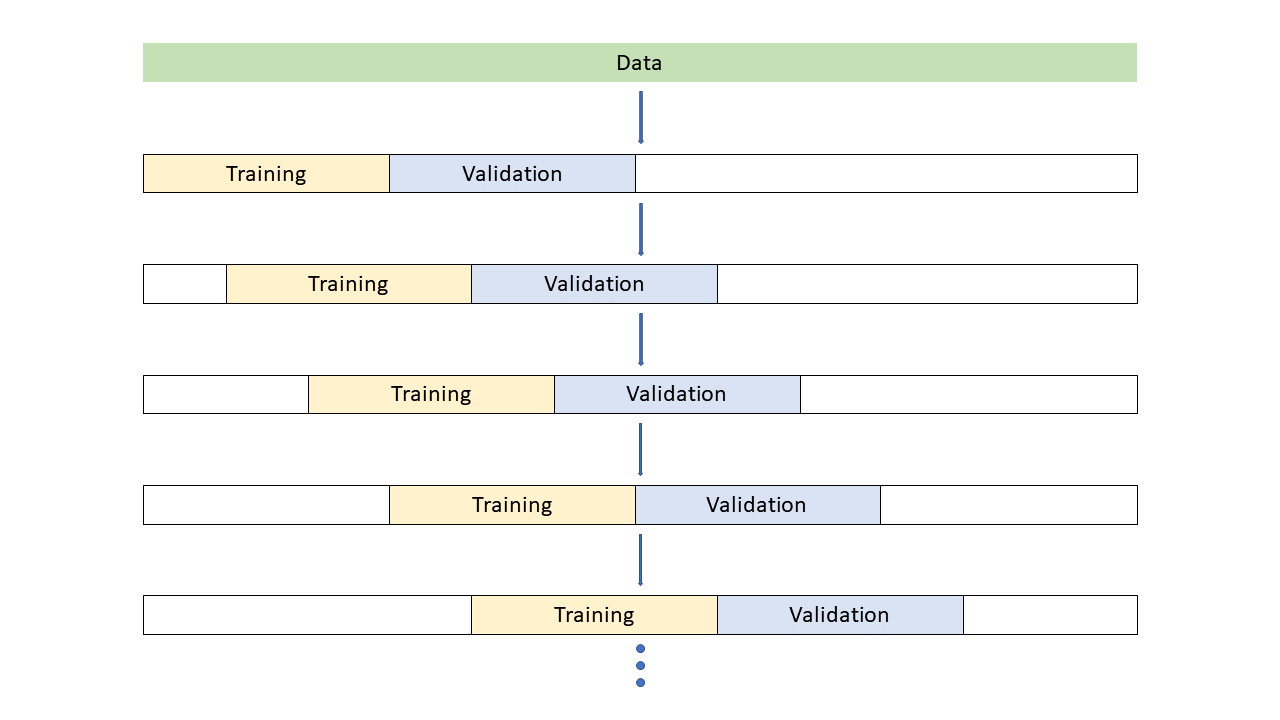}
\centering
\end{figure}

Figure 1 shows an example of it. Each block of the training set is divided into two parts at each iteration on the condition that the validation set is always ahead of the training split. This way, the natural order of observations is kept within each block. The number of iterations in the cross-validation setup depends on the size of each block. If it is small, the number of blocks and thus iterations increases. If I choose an overly small size of blocks, a certain block may have too few or even no data on recessions, which may result in sampling bias. The longest period between the start of a recession and the end of the next recession in US history prior to the Great Recession is approximately 12 years. That is why I let the size of each block or the length of the time series in each block be equal to $144\times2=288$ months or 24 years, ensuring that each training and validation set has data of at least one recession. The blocks overlap: The shortest period between two recessions from 1967 to 2006 is 1 year. Hence, I proceed with a rolling window scheme with the number of increments between successive rolling windows set to be 1 year. This ensures that scarce data on recessions are fully exploited.

Table 3 reports the ranges of hyperparameters used for grid search in the optimization process. The parameter space for each model is set to have the size of $10^3$.

\begin{table}[!ht]
    \caption{\\Tuning hyperparameters} 
    \begin{tabular}{l|c|r}
    \hline
    Model & Type & Range \\
    \hline
    LASSO & $\lambda$ & $10^{-5}$ to $10^2$ \\
    & & grid: 1000 \\
    \hline
    Ridge & $\lambda$ & $10^{-5}$ to $10^2$ \\
    & & grid: 1000 \\
    \hline
    \multirow{4}{*}{Elastic Net} & $\alpha$ & $\in(0,0.25,0.5,0.75,1)$ \\
    & & grid: 5 \\
    \cline{2-3}
    & $\lambda$ & $10^{-5}$ to $10^2$ \\
    & & grid: 200 \\
    \hline
    \end{tabular}\\
\end{table}

Once the models produce the probability predictions, they are translated into monthly zero-one indicators signaling the state of the economy, i.e., recession or boom, based on cutpoints. The traditional approach is to use a fixed threshold, which usually amounts to 0.5. In this case, a probability prediction greater than or equal to 0.5 is classified as a recession and vice versa. However, as the sample is unbalanced, choosing a 50\% threshold might not be optimal. \citet{berge2011} propose an optimal cutpoint between 0.3 and 0.6 based on the smoothed state probabilities estimated by \citet{chauvet2008}. \citet{ng2014} suggest different thresholds for different forecast horizons that range between 0.3 and 0.44. \citet{vrontos2021} choose a fixed threshold of 0.33 for classification and evaluation purposes.

I go one step further and internalize the threshold optimization: During the cross-validation process, I let the model compute the optimal cutpoint that minimizes the asymmetric misclassification costs of the validation set in each iteration. At the end of cross-validation, there are as many thresholds as the number of iterations. The mean of these thresholds represents the optimal threshold for out-of-sample forecasting. This optimal threshold is only valid for that vintage and varies over time. Misclassification costs are the costs that occur when recessions are classified as booms and booms as recessions. In the previous literature, there are some conflicting arguments about the costs of business cycles. \citet{lucas1987} argues that the welfare gains from reducing the volatility of aggregate consumption are negligible, which is why efforts to stabilize the economy are less important. \citet{clark1994} and \citet{barlevy2004}, on the other hand, provide empirical evidence that fluctuations reduce the growth rate of consumption, and business cycles are thus associated with a significant amount of costs. \citet{barro2006} and \citet{krebs2007} argue about the asymmetric costs of business cycles. They compute the welfare costs of events that are either the cause or the consequence of recessions and indicate that recessions are associated with higher costs. \citet{nevalsami2022} uses a cost-sensitive learning technique that is designed to address these asymmetric misclassification costs for the purpose of recession forecasting. I follow his approach and define the costs as equal to the class weights in the sample that are used for the weighted logit model. The cost of false negatives is thus set higher than the cost of false positives. The optimal threshold is then the cutpoint that minimizes the sum of costs with respect to false positives and false negatives.

The out-of-sample period begins in October 2005. For each month in the out-of-sample period, the vintage contains time series data of the NBER recession indicator and the predictors. This dataset is used to train the model and to forecast for the next period. Specifically, forecasting exercises are assumed to be conducted on the first day of every month during the out-of-sample forecasting period. The forecaster utilizes all available data up to the last day of the previous month to predict for that month and onwards, depending on the forecast horizon. Apart from data with a quarterly frequency, most of the data for a certain month are published on the last day of the next month at the latest. Hence, depending on the forecast horizon, there is at least a two-month lag between the month the latest available data refers to and the prediction month.    
\citet{stock1989} show that different macroeconomic indicators have different lagging properties. It means the length of this time lag differs between the leading macroeconomic indicators. That is why there is a deviation in the effectiveness of the predictive variables over different forecasting horizons. I exploit five distinct windows: nowcasting, immediate-term, short-term, medium-term, and long-term windows. I use a number of lags to increase the forecasting accuracy potentially. Specifically, nowcasting forecasts are computed using the predictive variables lagged by 2, 3, 6, and 12 months:

\begin{align*}
   \hat y_{t}=\hat f (\bm{x}_{t-2}, \bm{x}_{t-3}, \bm{x}_{t-6}, \bm{x}_{t-12}). 
\end{align*}
          
Similarly, forecasts for other forecast horizons are given by:
\begin{align*}
    \hat y_{t+1}&=\hat f (\bm{x}_{t-2}, \bm{x}_{t-5}, \bm{x}_{t-11}),\\
    \hat y_{t+3}&=\hat f (\bm{x}_{t-2}, \bm{x}_{t-3}, \bm{x}_{t-9}),\\
    \hat y_{t+6}&=\hat f (\bm{x}_{t-2}, \bm{x}_{t-6}), \\
    \hat y_{t+12}&=\hat f (\bm{x}_{t-2}).
\end{align*}

The latest available data, i.e., $\bm{x}_{t-2}$ with two months publications lags, are always used as predictors. Then, the predictive variables are lagged such that, depending on the forecast horizon, their lagging properties of the following magnitude are exploited: $l \in (0,1,3,6,12)$. They correspond to current-month, one-month-ahead, three-month-ahead, sixth-month-ahead, and twelve-month-ahead forecasting, respectively. Due to the two months of publication lags, the forecast horizon equals 2-, 3-, 5-, 8-, and 14-step-ahead forecasts. To illustrate this, suppose that I am in November 2006. In that month, I have data for September 2006 that was published at the end of October 2006. Now, I want to conduct nowcasting and forecast a recession for the current month. This means that I use the data from September 2006 to predict November 2006, resulting in a two-step-ahead forecast. Similarly, the other forecast horizons must be adjusted by two extra months or steps.  

The out-of-sample forecasts are obtained via an expanding window scheme. Based on the optimal hyperparameters obtained through the blocked cross-validation, a nowcast for November 2006 is computed. For the next iteration, the training and validation set is increased by one observation or month, and the model is re-estimated to produce the nowcast for December 2006. This procedure is repeated to generate out-of-sample forecasts for the testing period from November 2006 to October 2021. The same approach is adopted for the other forecast horizons.

\subsection{Performance evaluation}

The estimated models produce two types of forecasts: probability and point predictions. The models first generate probability predictions. The next step transforms these probabilities into binary point predictions using a specified threshold. While the second type may not always be necessary, it is included to ensure the forecast is more easily interpretable for the end users. The performance of both the probability and point predictions is evaluated using various statistical measures. Except for metrics related explicitly to probability predictions, the evaluation is based on the output of a contingency table, commonly known as a confusion matrix, as illustrated in Table 4. This matrix provides a framework for assessing the accuracy of the point predictions.

\begin{table}[!ht]
        \caption{Confusion Matrix}
    \begin{tabular}{c|c|c|c}
    \hline
        \multicolumn{2}{c|}{\multirow{2}{*}{}} & \multicolumn{2}{|c}{Predicted} \\
        \cline{3-4} 
        \multicolumn{2}{c|}{} & Positive & Negative \\
        \hline
        \multirow{2}{*}{Actual} & Positive & TP & FN \\
         & Negative & FP & TN \\
         \hline
    \end{tabular}\\
    \caption*{The table contains elements representing the count of observations belonging to each category. True positives (TP) and true negatives (TN) indicate the correct classification of positive and negative outcomes. False positives (FP) occur when the prediction is positive, but the actual value is negative. Conversely, false negatives (FN) arise when the prediction is negative, but the actual value is positive.} 
\end{table} 

The confusion matrix comprises elements representing the count of observations belonging to different categories. True positives (TP) and true negatives (TN) indicate the correct classification of positive and negative outcomes. False positives (FP) occur when the prediction is positive while the actual value is negative. Conversely, false negatives (FN) arise when the prediction is negative while the actual value is positive. By utilizing these counts, various performance evaluation measures are derived to assess the overall performance of the models.

\begin{table}[!ht]
    \caption{Performance evaluation metrics} 
    \resizebox{\textwidth}{!}{\begin{tabular}{c|l|c}
    \hline
    Type & Metric & Formula \\
    \hline
    Probability & Area under the ROC curve & $\int_0^1 ROC(c)\,dc$ \\  
    Prediction & Area under the PR curve & $\int_0^1 PR(c)\,dc$\\ 
    \hhline{=|=|=}
     & Sensitivity & $\frac{TP}{TP+FN}$\\  
     & Specificity & $\frac{TN}{FP+TN}$\\ 
     Point & Precision & $\frac{TP}{TP+FP}$ \\ 
     Prediction & Balanced accuracy & $\frac{Sensitivity+Specificity}{2}$\\  
     & Matthews correlation coefficient & \begin{tabular}{c}$\frac{TP\times TN-FP\times FN}{\sqrt{(TP+FP)(TP+FN)(TN+FP)(TN+FN)}}$\end{tabular}\\  
     & $F_1$-Score & $\frac{2}{\frac{1}{Sensitivity}+\frac{1}{Precision}}$\\  
    \hline
    \end{tabular}}
     \caption*{The table reports the metrics used for the performance evaluation of a forecasting model. They are divided into two groups, depending on which type of predictions they refer to.} 
\end{table}

Table 5 reports the list of the metrics for the performance evaluation. A carefully selected set of statistical metrics is used to assess the performance from different perspectives. The metrics, Area Under the Receiver Operating Characteristic Curve (AUROC) and Area Under the Precision-Recall Curve (AUPRC), are employed to measure the models' raw performance in the sense that these metrics do not require any threshold to convert probabilities into binary outcomes, making them suitable for evaluating the models' pure performance. The Receiver Operating Characteristics (ROC) curve displays the entire set of possible combinations of true positive rates $TPR(c)=\frac{TP(c)}{\#\:Actual\:Positive}$ and false positive rates $FPR(c)=\frac{FP(c)}{\#\:Actual\:Negative}$ for some cutpoint $c\in(0,1)$ that maps the predicted probability to a binary category. Similarly, the Precision Recall (PR) curve plots the complete set of possible combinations of precision and recall for $c\in(0,1)$, where recall is a synonym for sensitivity. The Area under the ROC and PR curves increases with the underlying metrics for a given cutpoint. By aggregating over the entire set of cutpoints, these curves deliver a framework to assess the pure predictive ability of a forecasting model. \citet{tharwat2021} provides a comprehensive overview of other metrics and discusses their strengths and weaknesses, particularly in imbalanced binary classification problems.

\section{Empirical results}

\renewcommand{\arraystretch}{0.74}
\begin{table}[!ht]
    \caption{\\Performance evaluation measures: A real-time assessment} 
    \resizebox{\textwidth}{!}{\begin{tabular}{lrrrrrrrr}
    \hline
    Method & AUROC & AUPRC & BAcc & MCC & $F_1$-Score & Sensitivity & Specificity & Precision \\
    \hline
    \multicolumn{9}{l}{Panel A: nowcasting setup (h=0)} \\
    \hline
    Logit & 0.853 & 0.365 & 0.816 & 0.507 & 0.556 & 0.750 & 0.881 & 0.441 \\
    wLogit & 0.845 & 0.367 & 0.816 & 0.507 & 0.556 & 0.750 & 0.881 & 0.441 \\
    LASSO & 0.868 & 0.387 & 0.725 & 0.396 & 0.468 & 0.550 & 0.900 & 0.407\\
    Ridge & 0.923 & 0.534 & 0.875 & 0.609 & 0.642 & 0.850 & 0.900 & 0.515 \\
    Elastic Net & 0.911 & 0.452 & 0.781 & 0.495 & 0.553 & 0.650 & 0.912 & 0.481\\
    \hline
    \multicolumn{9}{l}{Panel B: immediate-term setup (h=1)} \\
    \hline
    Logit & 0.671 &  0.243 & 0.669 & 0.265 & 0.357 & 0.500 & 0.838 & 0.278\\
    wLogit & 0.751 & 0.258 & 0.619 & 0.191 & 0.296 & 0.400 & 0.838 & 0.235 \\
    LASSO & 0.875 & 0.398 & 0.719 & 0.374 & 0.449 & 0.550 & 0.887 & 0.379\\
    Ridge & 0.931 & 0.555 & 0.894 & 0.693 & 0.723 & 0.850 & 0.938 & 0.630 \\
    Elastic Net & 0.893 & 0.398 & 0.819 & 0.518 & 0.566 & 0.750 & 0.887 & 0.455\\
    \hline
    \multicolumn{9}{l}{Panel C: short-term setup (h=3)} \\
    \hline
    Logit & 0.524 & 0.170 & 0.591 & 0.165 & 0.267 & 0.300 & 0.881 & 0.240\\
    wLogit & 0.655 & 0.216 & 0.656 & 0.236 & 0.333 & 0.500 & 0.812 & 0.250 \\
    LASSO & 0.895 & 0.380 & 0.797 & 0.494 & 0.549 & 0.700 & 0.894 & 0.452\\
    Ridge & 0.927 & 0.519 & 0.853 & 0.588 & 0.627 & 0.800 & 0.906 & 0.516 \\
    Elastic Net & 0.890 & 0.378 & 0.756 & 0.407 & 0.473 & 0.650 & 0.863 & 0.371\\
    \hline
    \multicolumn{9}{l}{Panel D: medium-term setup (h=6)} \\
    \hline
    Logit & 0.628 & 0.171 & 0.562 & 0.092 & 0.222 & 0.350 & 0.775 & 0.163 \\
    wLogit & 0.723 & 0.219 & 0.637 & 0.198 & 0.303 & 0.500 & 0.775 & 0.217 \\
    LASSO & 0.772 & 0.247 & 0.681 & 0.272 & 0.361 & 0.550 & 0.812 & 0.268\\
    Ridge & 0.907 & 0.416 & 0.819 & 0.518 & 0.566 & 0.750 & 0.887 & 0.455 \\
    Elastic Net & 0.828 & 0.305 & 0.728 & 0.362 & 0.436 & 0.600 & 0.856 & 0.343\\
    \hline
     \multicolumn{9}{l}{Panel E: long-term setup (h=12)} \\
    \hline
    Logit & 0.735 & 0.249 & 0.669 & 0.297 & 0.383 & 0.450 & 0.887 & 0.333 \\
    wLogit & 0.801 & 0.311 & 0.731 & 0.321 & 0.389 & 0.700 & 0.762 & 0.269 \\
    LASSO & 0.835 & 0.358 & 0.750 & 0.389 & 0.456 & 0.650 & 0.850 & 0.351\\
    Ridge & 0.816 & 0.385 & 0.709 & 0.344 & 0.423 & 0.550 & 0.869 & 0.344 \\
    Elastic Net & 0.826 & 0.301 & 0.697 & 0.309 & 0.393 & 0.550 & 0.844 & 0.306\\
    \hline
    \end{tabular}}
     \caption*{The table reports the performance evaluation measures of forecasts obtained by logit models, weighted logit models, and a series of penalized logit models over different time horizons for the out-of-sample period, November 2006 to October 2021: Panel (A) presents the nowcasts. Panel (B), (C), (D), and (E) display the 1-month-ahead forecasts, the 3-month-ahead forecasts, the 6-month-ahead forecasts, and the 12-month-ahead forecasts, respectively.}
\end{table}

This section presents the empirical findings based on the model specifications and data described above. Table 6 documents the models' out-of-sample forecast performance for each forecast horizon. The AUROC for a random classifier is fixed at 0.5. Compared to this classifier, all models show a significant improvement in prediction ability, except for the standard logit model for the short-term setup, where the difference is minimal. The overall results are consistent with recent studies, such as \citet{liu2016} and \citet{vrontos2021}. The AUPRC varies between 0.17 and 0.55. For a PR curve, the actual percentage of positive outcomes in the sample forms the lower bound. As mentioned above, the average percentage of months in a recession is approximately 20\%. Hence, the AUPRC for a random classifier can be thought to fluctuate around the value of 0.2. Bearing this in mind, the standard logit models again perform hardly better than random classifiers, or even worse in both the short-term and medium-term setup. However, other models show significant improvement in all setups.

For the nowcasting setup, there is little difference in predictive performance between two groups of models: the standard and the penalized logit models. In fact, the LASSO model exhibits similar performance compared to the standard logit models with regard to AUROC and AUPRC but performs worse in terms of other metrics related to point predictions. While both the LASSO and Elastic Net models do not seem to have an advantage over the standard logit models, the Ridge model outperforms the other models with a balanced accuracy of 87.5\%. Furthermore, it is able to classify 85\% of months in recession correctly while maintaining a precision value over 0.5. In the immediate-term setup, both the standard and weighted logit models perform significantly worse, with MCC value and $F_1$-Score approximately 50\% less than before. On the contrary, the series of penalized logit models show similar or slightly better performance. The Ridge model presents the highest performance across different measures such as balanced accuracy, MCC, and $F_1$-Score. In the short-term setup, the weighted logit model outperforms the standard one. Three penalized logit models have higher predictive performance than the other models. Again, the Ridge model is the best one among others with 80\% of correctly classified months in recession. For the medium-term setup, the pattern is similar. The weighted logit model outperforms the benchmark model but underperforms the penalized logit models. The Ridge model beats the other models with 75\% of true positives. In the long-term setup, the LASSO model outperforms the other models. It still manages to predict 65\% of the months in recession correctly.

\begin{figure}[!ht]
    \centering
    \begin{tabular}{c c}
        (A) &  (B) \\
        \includegraphics[width=.5\textwidth]{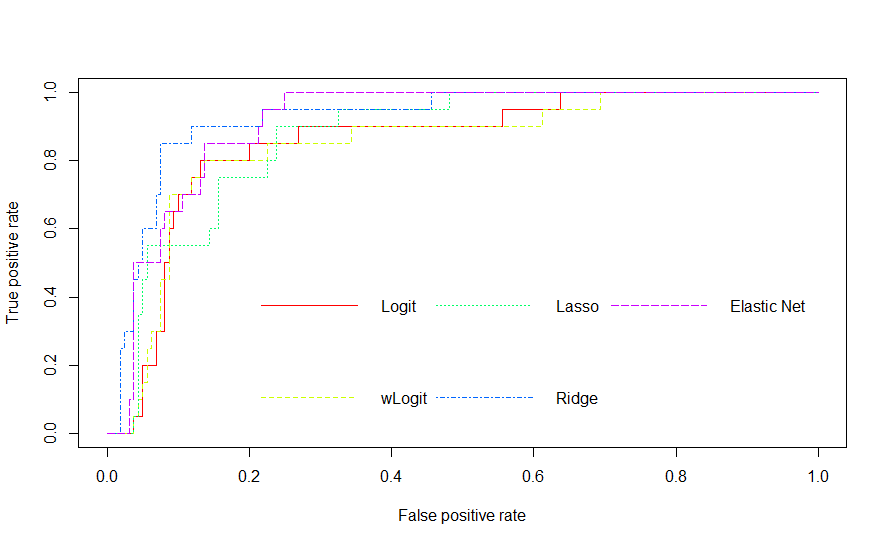} &  \includegraphics[width=.5\textwidth]{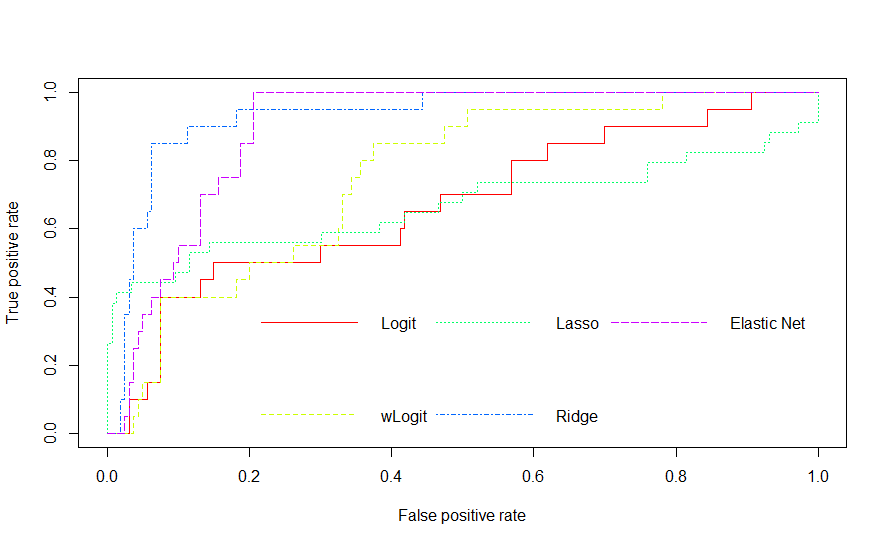} \\
        (C) & (D) \\
        \includegraphics[width=.5\textwidth]{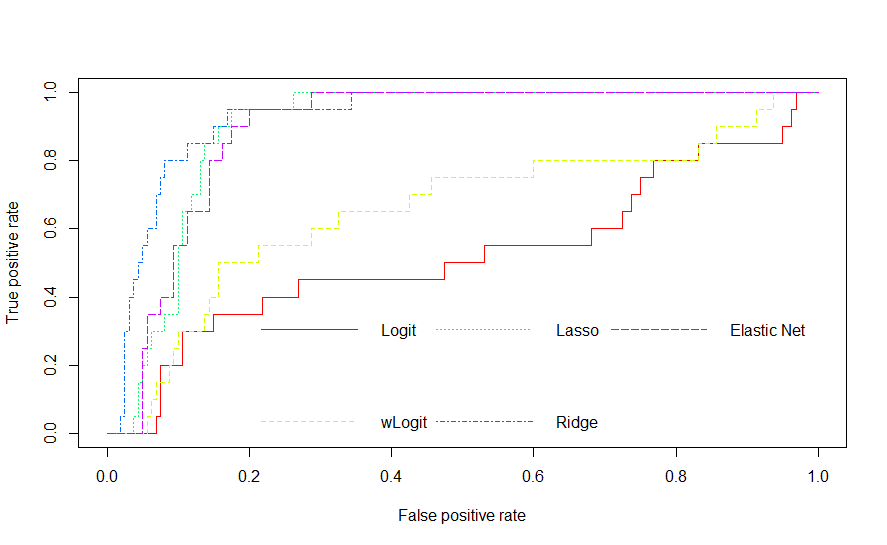} & \includegraphics[width=.5\textwidth]{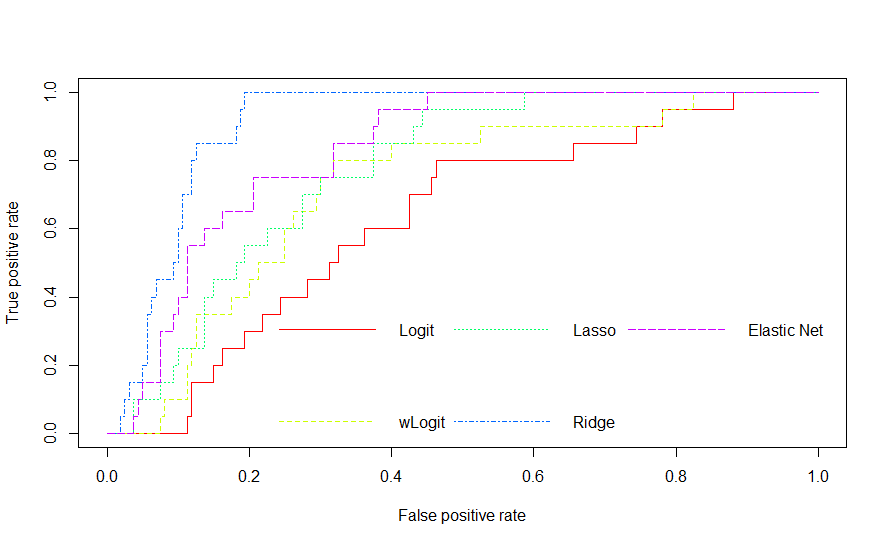} \\
        (E) & \\
        \includegraphics[width=.5\textwidth]{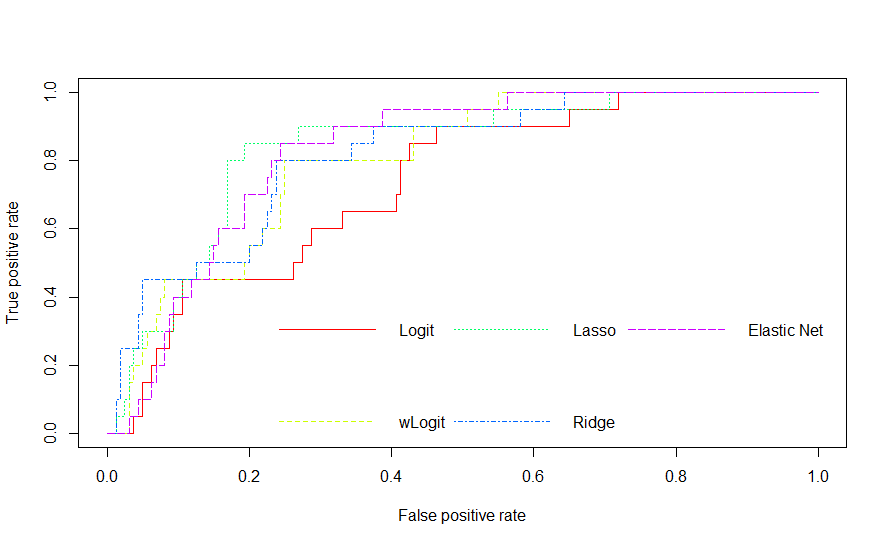} & \\
    \end{tabular}
    \caption{ROC curves for the out-of-sample period, November 2006 to October 2021: Panel (A) to (E) display the ROC curves based on the logit, weighted logit, the LASSO, the Ridge, and the Elastic Net model related to the nowcasting, the immediate-term (1-months-ahead), short-term (3-months-ahead), medium-term (6-months-ahead), and long-term (12-months-ahead) forecast horizon, respectively.}
\end{figure}

\begin{figure}[!ht]
    \centering
    \renewcommand{\arraystretch}{0.6}
    \begin{tabular}{c c}
        (A) &  (B) \\
        \includegraphics[width=.5\textwidth]{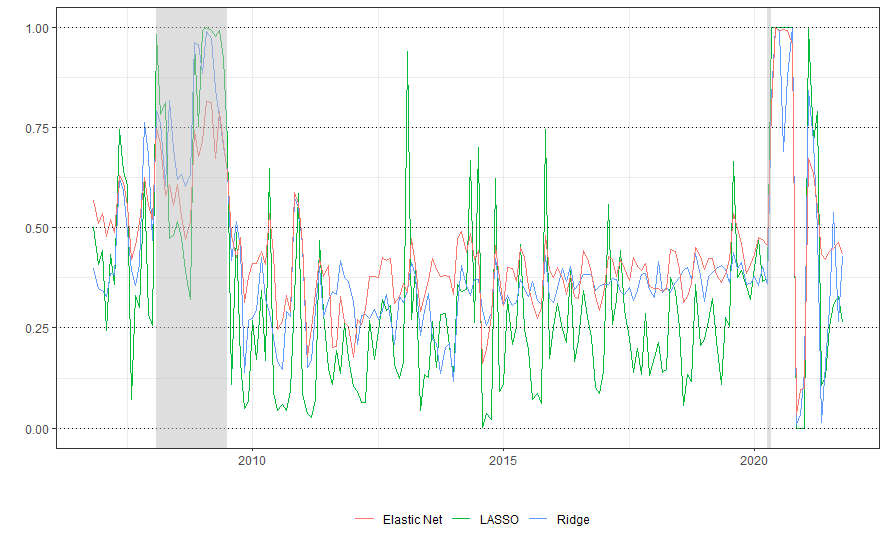} &  \includegraphics[width=.5\textwidth]{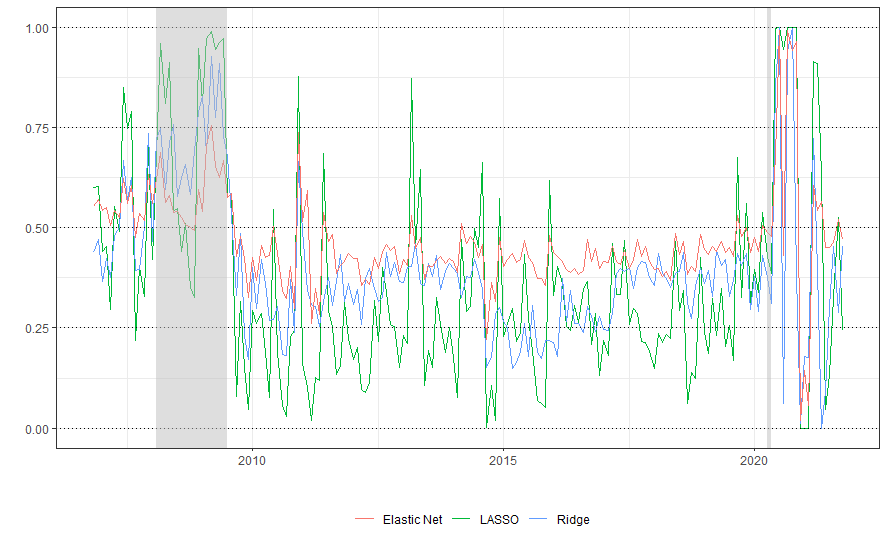} \\
        (C) & (D) \\
        \includegraphics[width=.5\textwidth]{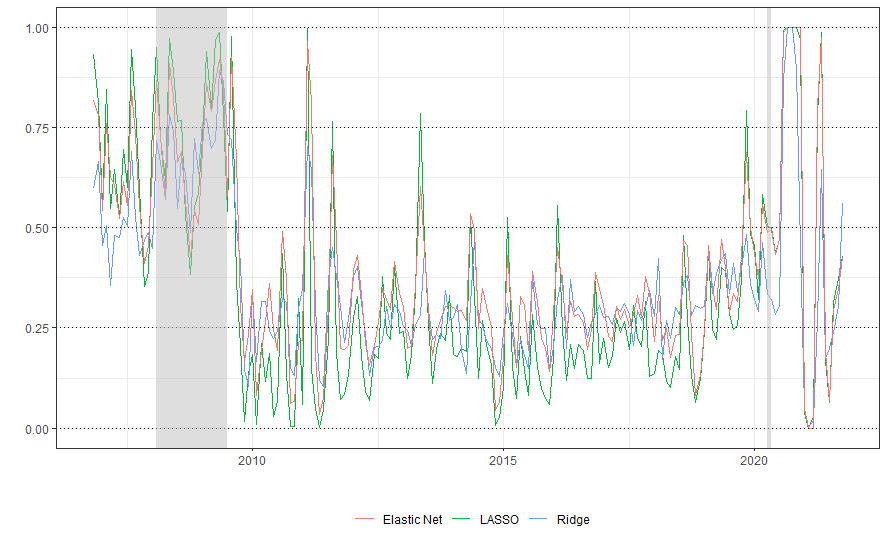} & \includegraphics[width=.5\textwidth]{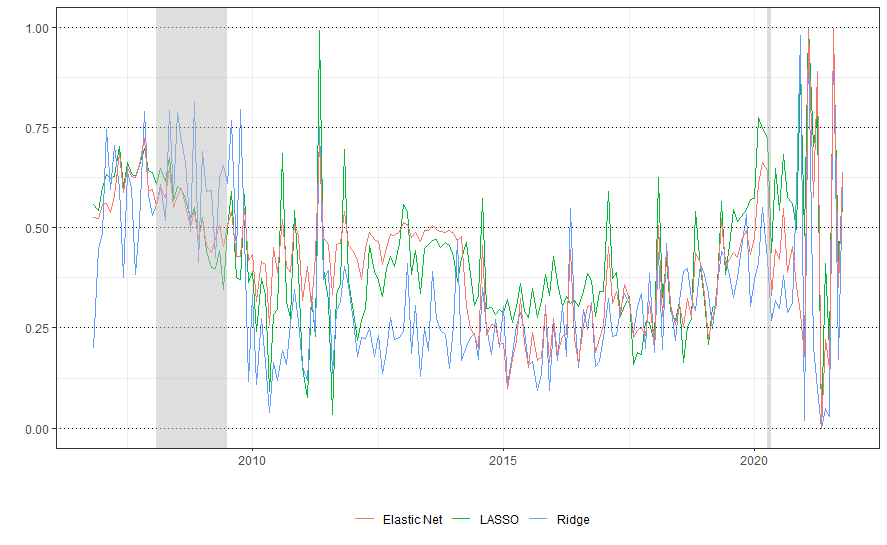} \\
        (E) & \\
        \includegraphics[width=.5\textwidth]{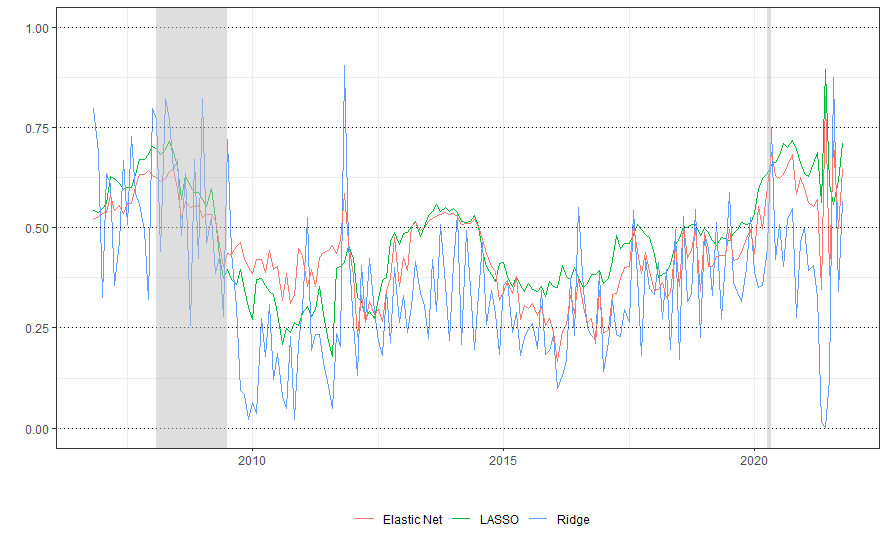} &  \\
    \end{tabular}
    \caption{Out-of-sample Recession Probabilities for the November 2006 to October 2021 period: Panel (A) to (E) display the predicted recession probabilities based on the LASSO, the Ridge, and the Elastic Net model related to the nowcasting, the immediate-term (1-month-ahead), short-term (3-months-ahead), medium-term (6-months-ahead), and long-term (12-months-ahead) forecast horizon, respectively. The grey shaded area depicts NBER recession months.}
\end{figure}

The ROC curves of five different models for five different forecast horizons are plotted in Figure 2. The difference in forecast performance between the standard logit model (red line) and the penalized logit models (blue and purple lines in particular) becomes evident in panels B, C, and D. Although the LASSO model (green line) does not outperform the standard logit models (red and yellow lines) in some of the setups, the Ridge model (blue line) dominates in almost all settings. The yellow lines, which represent the ROC curves of the weighted logit models, lie in almost all settings above the red line, supporting the different weighting schemes to account for class imbalance.

When the set of predictors is large, standard logit models often suffer from the problem of multicollinearity and complete separation. \citet{albert1984} define the latter as being the case if a predictor or a set of predictors correctly allocates all observations to their group in the training set. The forecasts are solely based on this set of predictors, which results in 100\% predicted probabilities. They prove that the maximum likelihood estimate of a logit model does not exist in the case of complete separation. \citet{mccullough2003} show that statistical software often outputs a parameter estimate of arbitrarily large magnitude with a very large standard error. This also happens when the dataset is characterized by severe multicollinearity. The weighted logit model only considers different weights for the observations belonging to different classes and is thus unaffected by the abovementioned problems. An approach to alleviate these problems with maximum likelihood estimation is regularization. The penalized logit models have the regularization terms in the maximum likelihood estimation and ensure that some parameter estimates are kept small or set to zero. Hence, the expectation is that at least one of these penalized logit models should exhibit higher predictive performance. The results meet this expectation as they outperform the logit models in all setups. Henceforth, the following analysis of the recessions focuses more on the three types of penalized logit models: the LASSO, the Ridge, and the Elastic Net model.

Figure 3 presents the out-of-sample forecasts of three penalized logit models in probabilities from November 2006 to October 2021 along with two shaded areas representing recessions. I exclude the other two variants because both logit models, without penalizing terms in the maximum likelihood estimation, sometimes produce results that are very close to either 1 or 0 due to the problem of complete separation. Hence, the graphs zigzag wildly making them hard to interpret. Furthermore, the penalized logit models outperform them in every setting. They are thus more appropriate for evaluating and comparing the predictability of the two recent recessions, the Great Recession and the Covid-19 recession. All panels in Figure 3 show similar patterns: The graphs start high, stay or get higher during the Great Recession, decrease after it, and fluctuate around the long-term average until they swing wildly after the Covid-19 recession. The models successfully signal economic and financial downturns during the Great Recession, but an exogenous event such as the Covid-19 recession remains largely unpredictable. The Ridge model in the immediate-term setup manages to predict up to 17 out of 18 recession months correctly during the Great Recession, while it fails to forecast any of the Covid-19 recession months. Depending on the forecast horizon, once the model learns from the impact of the Covid-19 pandemic on the economy, it struggles to interpret this sudden change and produce alternating forecasts.

Figure 3 alone does not provide any details about what drives the results. Hence, it is reasonable to scrutinize the development of the estimated parameters around the recessions. However, the predictor set is too large to examine every single variable. The LASSO model automatically selects variables by setting irrelevant variables' coefficient estimates to zero. Whenever the model is fit to produce the out-of-sample forecast, the variable or the set of variables included in the prediction is recorded. In this way, I find out how many times out of 180 out-of-sample forecasts a certain variable is included in the model. Table 7 presents the selected predictors for different forecast horizons with their number of lags and inclusion frequency in absolute numbers and percentages. The table reports only those variables whose inclusion frequency is higher than 80\%. This cutpoint is chosen because there is a significant drop in the inclusion rate below the 80\% level. 

\renewcommand{\arraystretch}{0.8}
\begin{table}[!ht]
    \caption{\\ LASSO model: Selected predictors} 
    \resizebox{\textwidth}{!}{\begin{tabular}{llll}
    \hline
    Variable & Lag & Inclusion frequency & Inclusion frequency \\
    & & (\# of times) & (\%) \\
    \hline
    \multicolumn{4}{l}{Panel A: nowcasting setup (h=0)} \\
    \hline
    Term spread - 10-year treasury yield minus 3-month treasury bill rate & 6 & 180 & 100\% \\
    Term spread - 10-year treasury yield minus 3-month treasury bill rate & 12 & 178 & 98.9\% \\
    S\&P 500 index & 3 & 178 & 98.9\%\\
    Real gross domestic product & 2 & 176 & 97.8\% \\
    Real gross domestic product & 6 & 176 & 97.8\% \\
    Private residential fixed investment & 3 & 167 & 92.8\% \\
    3-month treasury bill secondary market rate & 2 & 165 & 91.7\% \\
    Total nonfarm payroll - all employees & 2 & 164 & 91.1\% \\
    Private residential fixed investment & 12 & 161 & 89.4\% \\
    \hline
    \multicolumn{4}{l}{Panel B: immediate-term setup (h=1)} \\
    \hline
    Term spread - 10-year treasury yield minus 3-month treasury bill rate & 6 & 180 & 100\% \\
    Term spread - 10-year treasury yield minus 3-month treasury bill rate & 12 & 180 & 100\% \\
    Private residential fixed investment & 3 & 180 & 100\% \\
    Real gross domestic product & 6 & 174 & 96.7\% \\
    Real gross domestic product & 3 & 172 & 95.6\% \\
    Private residential fixed investment & 12 & 169 & 93.9\% \\
    Total nonfarm payroll - all employees & 3 & 150 & 83.3\% \\
    \hline
    \multicolumn{4}{l}{Panel C: short-term setup (h=3)} \\
    \hline
    Term spread - 10-year treasury yield minus 3-month treasury bill rate & 6 & 180 & 100\% \\
    Term spread - 10-year treasury yield minus 3-month treasury bill rate & 12 & 180 & 100\% \\
    Private residential fixed investment & 12 & 180 & 100\% \\
    Private residential fixed investment & 5 & 179 & 99.4\% \\
    Real gross domestic product & 5 & 173 & 96.1\% \\
    Real gross domestic product & 6 & 172 & 95.6\% \\
    S\&P 500 index & 6 & 149 & 82.8\%\\
    \hline
    \multicolumn{4}{l}{Panel D: medium-term setup (h=6)} \\
    \hline
    Term spread - 10-year treasury yield minus 3-month treasury bill rate & 8 & 180 & 100\% \\
    Term spread - 10-year treasury yield minus 3-month treasury bill rate & 12 & 180 & 100\% \\
    Private residential fixed investment & 8 & 180 & 100\% \\
    \hline
    \multicolumn{4}{l}{Panel E: long-term setup (h=12)} \\
    \hline
    Term spread - 10-year treasury yield minus 3-month treasury bill rate & 14 & 180 & 100\% \\
    \hline
    \end{tabular}}
    \caption*{The table reports the most relevant predictive variables and their underlying frequency of inclusion obtained by the LASSO model regarding the long-term horizon (12-months-ahead) forecasts for the out-of-sample period (2006:11-2021:10).}
\end{table}

Apart from the long-term setup where there is only one variable, the variables are ranked based on their frequency of inclusion. To begin with, in the long-term setup, there is only one predictor, the term spread, that is included more than 80\% in the model to produce 12-month-ahead forecasts. In fact, it is never left out in the estimation. Surprisingly, there are only two variables that are ever used in the long-term setup: One of them is the term spread, and the other one is private residential fixed investment. However, the rate of inclusion of the latter amounts to 6\% only. The other 23 variables are completely left out by the model and found to have no long-term predictive ability. It shows a similar pattern in the medium-term setup. Again, the term spread with different lags is included in every model that is estimated to produce out-of-sample forecasts. Private residential fixed investment is another variable used in the model with 100\% inclusion frequency. In the short-term setup, the list of variables with a rate of inclusion above 80\% is extended to account for a real gross domestic product as well as the S\&P 500 index. As for the immediate-term setup, the list remains largely the same except for the S\&P 500 index replaced by total nonfarm payroll. Finally, in the nowcasting setup, the list is extended by the 3-month treasury bill. All in all, the results show that the models utilize only a few selected variables across different forecast horizons. Only 7 out of 25 variables at maximum are used as predictors in more than 80\% of the models. As the forecast horizon increases, the list shrinks, and one variable shows up in every list with 100\% inclusion frequency, the term spread. In particular, the term spread is considered one of the most important predictors in the existing literature. \citet{estrella1998}, for instance, present empirical results that support the inclusion of the term spread in the simple logit model regardless of forecast horizons. \citet{vrontos2021} include the term spread or yield curve as the sole predictive variable. In fact, they base their decision on the observation that there had been only a single false signal in the past 50 years, and the curve inverted ahead of each of the last seven recessions.

\renewcommand{\arraystretch}{0.8}
\begin{table}[!ht]
    \caption{\\Weighted logit models with LASSO selected variables} 
    \resizebox{\textwidth}{!}{\begin{tabular}{lrrrrrrrr}
    \hline
   Method & AUROC & AUPRC & BAcc & MCC & $F_1$-Score & Sensitivity & Specificity & Precision \\
    \hline
    \multicolumn{9}{l}{Panel A: nowcasting setup (h=0)} \\
    \hline
    All predictors & 0.845 & 0.367 & 0.816 & 0.507 & 0.556 & 0.750 & 0.881 & 0.441 \\
    Selected predictors & 0.878 & 0.394 & 0.738 & 0.390 & 0.462 & 0.600 & 0.875 & 0.375 \\
    \hline
    \multicolumn{9}{l}{Panel B: immediate-term setup (h=1)} \\
    \hline
    All predictors & 0.751 & 0.258 & 0.619 & 0.191 & 0.296 & 0.400 & 0.838 & 0.235 \\
    Selected predictors & 0.797 & 0.317 & 0.700 & 0.318 & 0.400 & 0.550 & 0.850 & 0.314 \\
    \hline
    \multicolumn{9}{l}{Panel C: short-term setup (h=3)} \\
    \hline
    All predictors & 0.655 & 0.216 & 0.656 & 0.236 & 0.333 & 0.500 & 0.812 & 0.250\\
    Selected predictors & 0.878 & 0.389 & 0.725 & 0.354 & 0.429 & 0.600 & 0.850 & 0.333 \\
    \hline
    \multicolumn{9}{l}{Panel D: medium-term setup (h=6)} \\
    \hline
    All predictors & 0.723 & 0.219 & 0.637 & 0.198 & 0.303 & 0.500 & 0.775 & 0.217\\
    Selected predictors & 0.792 & 0.278 & 0.762 & 0.366 & 0.423 & 0.750 & 0.775 & 0.294 \\
    \hline
    \multicolumn{9}{l}{Panel E: long-term setup (h=12)} \\
    \hline
    All predictors & 0.801 & 0.311 & 0.731 & 0.321 & 0.389 & 0.700 & 0.762 & 0.269\\
    Selected predictors & 0.823 & 0.347 & 0.719 & 0.337 & 0.414 & 0.600 & 0.838 & 0.316 \\
    \hline
    \end{tabular}}
    \caption*{The table reports the performance evaluation measures of forecasts obtained by weighted logit models, using all predictors as well as selected variables according to LASSO variable selection criteria over different time horizons for the out-of-sample period, November 2006 to October 2021: Panel (A) presents the nowcasts. Panel (B), (C), (D), and (E) display the 1-month-ahead forecasts, 3-month-ahead forecasts, the 6-month-ahead forecasts, and the 12-month-ahead forecasts, respectively.}
\end{table}

Table 8 reports the forecast performance of the weighted logit models with selected predictors listed in Table 7. This performance is then compared to the performance of the weighted logit models with all predictors. Except for the nowcasting setup, the weighted logit models with selected predictors exhibit superior forecasting ability compared to the weighted logit models with all predictors. For the threshold-invariant measures, AUROC and AUPRC, the models with selected predictors outperform the models with all predictors in all settings, even in the nowcasting setup. This result empirically demonstrates the potential existence of multicollinearity and the complete separation problem in the dataset, further strengthening the argument to treat the LASSO-selected variables as the main drivers of the recessions.

Figure 4 reports the development of the five LASSO-selected variables over time along with their regression coefficients and the predicted recession probabilities for the immediate-term forecast horizon. These selected variables are listed in the panel A of Table 7. In cases where there are three variables with the same name but different lags, the variable with the shorter lag is chosen, as they show better or at least equal inclusion frequency. The 3-month treasury bill rate is excluded in the analysis as it does not cover at least 80\% of the recession periods. The graphs in the second row display the development of the standardized variables for the time being predicted. Specifically, the data point plotted at time $t$ is from time $t-l$, where $l$ denotes the lag of the variable that is used to predict for time $t$. It is the same for the coefficients of the variables. This enables direct time matching with the predicted recession probabilities in the first row to determine which of these five variables contributes to the change in recession probabilities during the two recession periods and whether the models capture the predictability of the recessions in their coefficients.

Several conspicuous features need to be pointed out in Figure 4. First, it can be noticed that the first three variables, real GDP, total nonfarm payroll, and private residential fixed investment, do not vary much during the two recession periods. However, there are significant changes in the magnitude and direction of their coefficients, at least for the period of the Great Recession. It shows that the models can capture the effects of these variables on the recession probability during the Great Recession. For the period before and during the COVID-19 recession, there has been hardly any change for almost a year. They drop sharply only afterward. However, they must be considered together with the wild changes of the variables themselves.

\begin{landscape}
\begin{figure}[hb!]
    \centering
    \renewcommand{\arraystretch}{0.1}
    \includegraphics[width=\linewidth]{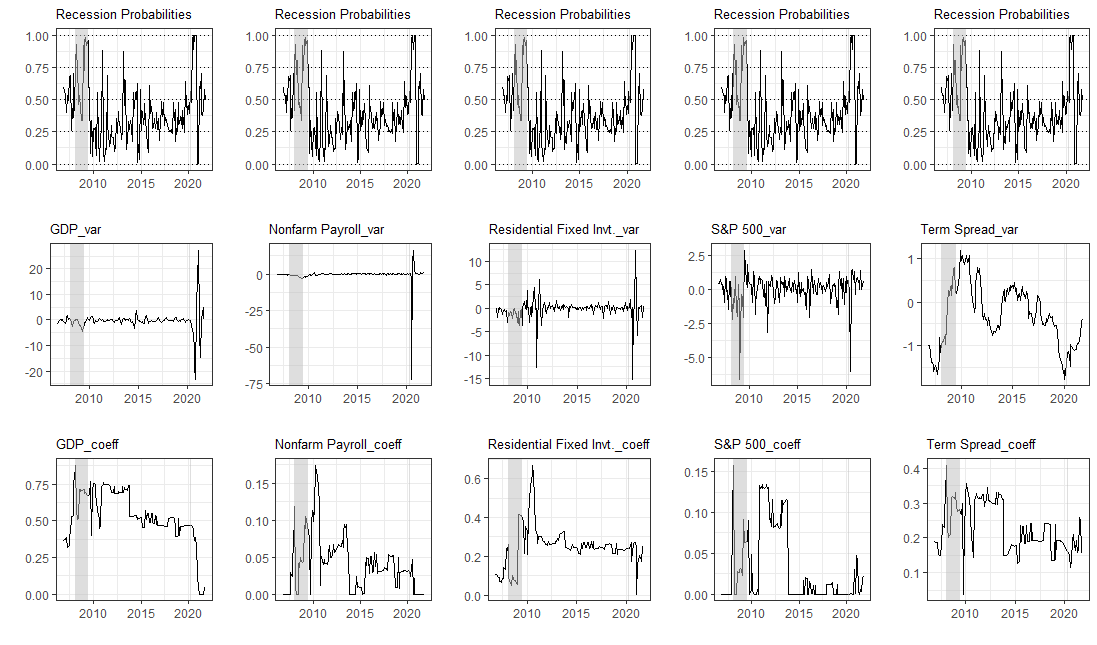}
    \caption{Out-of-sample recession probabilities and standardized values and coefficients of five LASSO selected variables for the immediate-term (1-month-ahead) forecast horizon.}
\end{figure}
\end{landscape}%

The complexity is evident in the second quarter of 2020 when the impact of the economic shutdown in March 2020 begins to materialize. The models attempt to adapt to these sudden changes in variables by suppressing the coefficients towards zero but fail to absorb them fully, leading to a sharp increase and following a sharp decrease in the predicted recession probabilities. The S\&P 500 index and the term spread in the fourth and fifth columns of Figure 4, respectively, can be characterized by dramatic changes in both the variables and the coefficients during the Great Recession. The S\&P 500 index plunged approximately 17\% for October 2008, often cited as the month in the depth of the Great Recession after Lehman's failure in mid-September 2008. Its coefficients are suppressed to zero during the first few months of the Great Recession but increase gradually afterward. Both the values and the coefficients for S\&P 500 contribute to high but fluctuating recession probabilities during the Great Recession. During the Covid-19 recession period, the S\&P 500 exhibits a huge decline in March 2020, attributed to the outbreak of the Covid-19 pandemic and the following economic shutdown. The coefficients are suppressed to zero around the Covid-19 recession period. This variable fails to deliver any information about the recession prior to this period.

The term spread turns negative before the Great Recession and increases sharply during that recession period. Afterward, it decreases gradually until it reaches a magnitude similar to that prior to the Great Recession. It increases during and after the Covid-19 recession. The coefficients also increase and remain high during the Great Recession. However, they decrease and stay low around the Covid-19 recession. The variable itself reacts to macroeconomic changes and indicates the need for the economy to stabilize prior to recessions.

However, the two recessions are of a different nature. The Great Recession was induced by the collapse of the financial system and was thus of a financial nature, similar to other recessions from 1967 that were often caused by high inflation and tightening monetary policy. The model recognizes this kind of recession and reacts to changes in the term spread by adjusting the coefficients. Prior to the Covid-19 recession, the economy may have been in an unstable state, but the recession was triggered by the Covid-19 pandemic and the subsequent economic shutdown. It is of a nature that cannot be predicted a priori, and moreover, the dataset does not contain such a case. Hence, the model reacts to the negative term spread in a different way, by reducing the effect of that variable. It becomes evident that major macroeconomic and financial indicators can predict a recession of an economic and financial nature. However, they cannot forecast a recession caused by an exogenous event such as a pandemic.

\section{Robustness checks}

This section introduces two alternative strategies designed to reduce the systematic error that arises when the exact time of the start and the end of a recession is announced by the NBER with considerable delay. In this setting, misleading data about recessions is fed into the model, potentially leading to considerable bias in the estimates.

The first approach, similar to \citet{hwang2019}, involves halting the updating of the model coefficients once the start of a recession is announced. This means that the model no longer has to deal with erroneous data. I illustrate the procedures using the example of the Great Recession. Table 2 indicates that the start of the Great Recession was announced in December 2008, retroactively declared to be in January 2008. According to the procedure described above, the model is updated each month until December 2008 assuming there is no recession. Once the start of the recession is announced in January 2009, all the zeros of the recession indicator between January 2008 and November 2008 are changed to ones. The model is then updated for the last time before the end of the recession is announced. Consequently, until September 2010 (according to Table 2), I do not update the model and continue producing out-of-sample forecasts based on new data. In October 2010, I update the recession indicator based on the announcement, setting it to 1 for the period between December 2008 and June 2009, and to 0 for the period between July 2009 and August 2010. This procedure is repeated for NBER announcements related to the Covid-19 recession.

\renewcommand{\arraystretch}{0.74}
\begin{table}[!ht]
    \caption{\\Performance evaluation measures: Alternative strategy} 
    \resizebox{\textwidth}{!}{\begin{tabular}{lrrrrrrrr}
    \hline
     Method & AUROC & AUPRC & BAcc & MCC & $F_1$-Score & Sensitivity & Specificity & Precision \\
    \hline
    \multicolumn{9}{l}{Panel A: nowcasting setup (h=0)} \\
    \hline
    Logit & 0.807 & 0.368 & 0.778 & 0.482 & 0.542 & 0.650 & 0.906 & 0.464 \\
    wLogit & 0.825 & 0.370 & 0.750 & 0.434 & 0.500 & 0.600 & 0.900 & 0.429 \\
    LASSO & 0.869 & 0.368 & 0.841 & 0.498 & 0.531 & 0.850 & 0.831 & 0.386\\
    Ridge & 0.910 & 0.496 & 0.878 & 0.622 & 0.654 & 0.850 & 0.906 & 0.531 \\
    Elastic Net & 0.911 & 0.452 & 0.781 & 0.495 & 0.553 & 0.650 & 0.912 & 0.481\\
    \hline
    \multicolumn{9}{l}{Panel B: immediate-term setup (h=1)} \\
    \hline
    Logit & 0.784 &  0.320 & 0.706 & 0.381 & 0.455 & 0.500 & 0.912 & 0.417\\
    wLogit & 0.830 & 0.336 & 0.753 & 0.367 & 0.431 & 0.700 & 0.806 & 0.311 \\
    LASSO & 0.887 & 0.408 & 0.756 & 0.458 & 0.522 & 0.600 & 0.912 & 0.462\\
    Ridge & 0.931 & 0.548 & 0.894 & 0.693 & 0.723 & 0.850 & 0.938 & 0.630 \\
    Elastic Net & 0.900 & 0.425 & 0.819 & 0.518 & 0.566 & 0.750 & 0.887 & 0.455\\
    \hline
    \multicolumn{9}{l}{Panel C: short-term setup (h=3)} \\
    \hline
    Logit & 0.696 & 0.213 & 0.625 & 0.205 & 0.308 & 0.400 & 0.850 & 0.250\\
    wLogit & 0.767 & 0.256 & 0.756 & 0.330 & 0.374 & 0.850 & 0.662 & 0.239 \\
    LASSO & 0.897 & 0.384 & 0.794 & 0.483 & 0.538 & 0.700 & 0.887 & 0.438\\
    Ridge & 0.918 & 0.502 & 0.881 & 0.635 & 0.667 & 0.850 & 0.912 & 0.548 \\
    Elastic Net & 0.889 & 0.384 & 0.781 & 0.442 & 0.500 & 0.700 & 0.863 & 0.389\\
    \hline
    \multicolumn{9}{l}{Panel D: medium-term setup (h=6)} \\
    \hline
    Logit & 0.583 & 0.175 & 0.562 & 0.100 & 0.222 & 0.300 & 0.825 & 0.176 \\
    wLogit & 0.685 & 0.214 & 0.722 & 0.284 & 0.344 & 0.800 & 0.644 & 0.219 \\
    LASSO & 0.823 & 0.301 & 0.738 & 0.356 & 0.426 & 0.650 & 0.825 & 0.317\\
    Ridge & 0.907 & 0.412 & 0.819 & 0.518 & 0.566 & 0.750 & 0.887 & 0.455 \\
    Elastic Net & 0.859 & 0.333 & 0.775 & 0.424 & 0.483 & 0.700 & 0.850 & 0.368\\
    \hline
     \multicolumn{9}{l}{Panel E: long-term setup (h=12)} \\
    \hline
    Logit & 0.721 & 0.273 & 0.675 & 0.258 & 0.349 & 0.550 & 0.800 & 0.256 \\
    wLogit & 0.797 & 0.357 & 0.703 & 0.299 & 0.381 & 0.600 & 0.806 & 0.279 \\
    LASSO & 0.856 & 0.450 & 0.719 & 0.337 & 0.414 & 0.600 & 0.838 & 0.316\\
    Ridge & 0.811 & 0.404 & 0.713 & 0.354 & 0.431 & 0.550 & 0.875 & 0.355 \\
    Elastic Net & 0.864 & 0.414 & 0.744 & 0.411 & 0.480 & 0.600 & 0.887 & 0.400\\
    \hline
    \end{tabular}}
     \caption*{The table reports the performance evaluation measures of forecasts obtained by logit models, weighted logit models, and a series of penalized logit models over different time horizons for the out-of-sample period, November 2006 to October 2021: Panel (A) presents the nowcasts. Panel (B), (C), (D), and (E) display the 1-month-ahead forecasts, the 3-month-ahead forecasts, the 6-month-ahead forecasts, and the 12-month-ahead forecasts, respectively.}
\end{table}

Table 9 presents the performance evaluation measures of this alternative strategy. In the nowcasting setup, this strategy outperforms the approach in the previous section in terms of classification metrics but underperforms it regarding both probability prediction metrics AUROC and AUPRC. However, the difference is minimal. Similar results are observed in the immediate-term setup. As the forecast horizon increases, the alternative strategy performs slightly better than the normal approach, although the difference remains negligible. Overall, Table 9 provides evidence that the main result in the previous section is robust against the systematic delay of NBER announcements.

The second approach considers the use of a completely different recession indicator that is available every month with less publication lag and is highly correlated with the NBER recession indicator. \citet{li2021} use a dynamic factor model to extract a common factor from four monthly coincident indicators: non-farm payroll, industrial production index, real personal income excluding current transfer receipts, and real manufacturing and trade industries sales. The NBER highlights these series in their decision on dating turning points. \citet{chauvet2008} use the same series to fit a dynamic factor Markov switching model to identify new turning points in real time. I conduct PCA on these variables to extract a common factor using only the first principal component. I apply the Bry and Boschan algorithm from \citet{bry1971} to this first principal component series to obtain new business cycle dates. These dates are then used to convert this series into a binary variable that can be interpreted as a recession indicator. This newly built recession indicator fulfills the requirements mentioned above. First, it shows, on average, a phi coefficient value of 0.92, indicating a very high association with the NBER recession indicator. The phi coefficient, or Matthews correlation coefficient, is essentially the Pearson correlation coefficient applied to two binary variables. Second, it is available every month with considerably less publication lag. For instance, it manages to identify the start and end of the Great Recession perfectly with only a 6-month delay for both the start and end of the recession. Compared to the publication lag reported in Table 2, this represents a significant improvement. In particular, the announcement delay for the end of the Great Recession is reduced by 9 months.  

\renewcommand{\arraystretch}{0.74}
\begin{table}[!ht]
    \caption{\\Performance evaluation measures: Alternative recession indicator} 
    \resizebox{\textwidth}{!}{\begin{tabular}{lrrrrrrrr}
    \hline
     Method & AUROC & AUPRC & BAcc & MCC & $F_1$-Score & Sensitivity & Specificity & Precision \\
    \hline
    \multicolumn{9}{l}{Panel A: nowcasting setup (h=0)} \\
    \hline
    Logit & 0.797 & 0.730 & 0.782 & 0.625 & 0.687 & 0.605 & 0.958 & 0.793\\
    wLogit & 0.811 & 0.762 & 0.778 & 0.609 & 0.676 & 0.605 & 0.951 & 0.767 \\
    LASSO & 0.835 & 0.754 & 0.741 & 0.521 & 0.609 & 0.553 & 0.930 & 0.677\\
    Ridge & 0.812 & 0.786 & 0.820 & 0.646 & 0.720 & 0.711 & 0.930 & 0.730 \\
    Elastic Net & 0.816 & 0.769 & 0.803 & 0.613 & 0.693 & 0.684 & 0.923 & 0.703\\
    \hline
    \multicolumn{9}{l}{Panel B: immediate-term setup (h=1)} \\
    \hline
    Logit & 0.749 &  0.659 & 0.731 & 0.514 & 0.597 & 0.526 & 0.937 & 0.690\\
    wLogit & 0.772 & 0.679 & 0.745 & 0.536 & 0.618 & 0.553 & 0.937 & 0.700 \\
    LASSO & 0.809 & 0.713 & 0.753 & 0.512 & 0.613 & 0.605 & 0.901 & 0.622\\
    Ridge & 0.757 & 0.696 & 0.754 & 0.543 & 0.629 & 0.579 & 0.930 & 0.688 \\
    Elastic Net & 0.863 & 0.767 & 0.799 & 0.572 & 0.667 & 0.711 & 0.887 & 0.628\\
    \hline
    \multicolumn{9}{l}{Panel C: short-term setup (h=3)} \\
    \hline
    Logit & 0.696 & 0.550 & 0.712 & 0.409 & 0.522 & 0.545 & 0.878 & 0.500\\
    wLogit & 0.741 & 0.586 & 0.727 & 0.434 & 0.543 & 0.576 & 0.878 & 0.514 \\
    LASSO & 0.852 & 0.645 & 0.782 & 0.540 & 0.642 & 0.684 & 0.880 & 0.605\\
    Ridge & 0.743 & 0.619 & 0.760 & 0.537 & 0.630 & 0.605 & 0.915 & 0.657 \\
    Elastic Net & 0.825 & 0.653 & 0.775 & 0.519 & 0.627 & 0.684 & 0.866 & 0.578\\
    \hline
    \multicolumn{9}{l}{Panel D: medium-term setup (h=6)} \\
    \hline
    Logit & 0.692 & 0.459 & 0.683 & 0.343 & 0.472 & 0.515 & 0.850 & 0.436 \\
    wLogit & 0.714 & 0.475 & 0.687 & 0.339 & 0.494 & 0.579 & 0.796 & 0.431 \\
    LASSO & 0.802 & 0.500 & 0.734 & 0.421 & 0.556 & 0.658 & 0.810 & 0.481\\
    Ridge & 0.729 & 0.524 & 0.730 & 0.455 & 0.571 & 0.579 & 0.880 & 0.564 \\
    Elastic Net & 0.761 & 0.543 & 0.709 & 0.411 & 0.538 & 0.553 & 0.866 & 0.525\\
    \hline
     \multicolumn{9}{l}{Panel E: long-term setup (h=12)} \\
    \hline
    Logit & 0.810 & 0.549 & 0.757 & 0.455 & 0.581 & 0.711 & 0.803 & 0.491 \\
    wLogit & 0.861 & 0.586 & 0.759 & 0.451 & 0.577 & 0.737 & 0.782 & 0.475 \\
    LASSO & 0.953 & 0.891 & 0.918 & 0.777 & 0.824 & 0.921 & 0.915 & 0.745\\
    Ridge & 0.777 & 0.505 & 0.712 & 0.403 & 0.537 & 0.579 & 0.845 & 0.500 \\
    Elastic Net & 0.949 & 0.863 & 0.879 & 0.719 & 0.780 & 0.842 & 0.915 & 0.727\\
    \hline
    \end{tabular}}
     \caption*{The table reports the performance evaluation measures of forecasts obtained by logit models, weighted logit models, and a series of penalized logit models over different time horizons for the out-of-sample period, November 2006 to October 2021: Panel (A) presents the nowcasts. Panel (B), (C), (D), and (E) display the 1-month-ahead forecasts, the 3-month-ahead forecasts, the 6-month-ahead forecasts, and the 12-month-ahead forecasts, respectively.}
\end{table}

Table 10 reports the results for this new recession indicator. The major difference is that AUPRC is significantly higher than before. This doesn't necessarily lead to higher performance in the point predictions of the penalized logit models. Still, it does make the standard and the weighted logit models perform notably better, narrowing down the performance spread. Surprisingly, the model performs better in the long-term setup than in the nowcasting or immediate-term setup. The LASSO model performs extraordinarily well with a balanced accuracy of 91.8\%. Once again, the variable selection procedure reveals the main driver of this result as the term spread. However, it is sensitive to the change in the validation method. For instance, when using an expanding window scheme in the cross-validation process, the performance evaluation measures tend to be slightly higher than for the standard and the weighted logit models. Both methods achieve, on average, higher performance than in the previous section and can thus be used in addition to the original method to strengthen its predictability.

\section{Conclusion}

This study examines the real-time predictability of the Great Recession and the Covid-19 recession compared to each other. Five different models, the logit model, weighted logit model, LASSO model, Ridge model, and Elastic Net model, are estimated using vintage data of 25 macroeconomic and financial market variables from February 1967 to October 2021. Out-of-sample forecasts are generated from November 2006 to October 2021 and evaluated by a set of metrics that measure performance evaluation for both probability and point predictions. Based on these measures, the penalized logit models are chosen for in-depth analysis of the two recessions. I examine the set of predictors chosen through the automatic variable selection of LASSO models to identify their different behaviors before the recessions. The publication lags of NBER business cycle dating pose a major concern, as misleading data train the models until the correct dates are announced. I exploit two other specifications of the baseline method to conduct robustness checks.

The main findings of this empirical study can be summarized in three points. Firstly, the Great Recession is largely predictable in real time, whereas the Covid-19 recession remains unpredictable due to the Covid-19 pandemic being a genuine exogenous event. A detailed analysis of the predictors driving these results reveals that they exhibit similar patterns before the Great Recession, remain calm before the Covid-19 recession, and swing wildly afterward due to the impact of the Covid-19 pandemic. Secondly, the penalized logit model, the Ridge model, outperforms the standard logit model, even after accounting for different class weights. These models effectively address issues like multicollinearity and complete separation by introducing a penalty term in the likelihood function that suppresses coefficients of highly correlated or less important variables. This enables them to generate more robust real-time forecasts and to deliver the set of most important variables through the automatic variable selection of the LASSO model. Thirdly, the study reaffirms the role of the term spread as the most important recession indicator. It is the only variable never left out of the LASSO model regardless of the forecast horizons and is the sole predictor in the long-term setup. 

\clearpage
\bibliographystyle{elsarticle-harv} 
\bibliography{cas-refs}





\end{document}